\documentclass[aps,prd,showpacs,nofootinbib,floats,floatfix,preprintnumbers,groupedaddress,twocolumn]{revtex4}

\usepackage{bm}
\usepackage{latexsym}
\usepackage{dcolumn}
\usepackage{amsmath,amsfonts,amssymb}
\usepackage{graphicx,epsfig}
\usepackage{amsmath}
\usepackage{fancyhdr}
\usepackage{hyperref}
\usepackage{graphicx,epstopdf}
\hypersetup{
	colorlinks   = true, 
	urlcolor     = blue, 
	linkcolor    = blue, 
	citecolor   = red 
}

\def\no{\nonumber}
\def\a{\alpha}
\def\b{\beta}
\def\g{\gamma}
\def\d{\delta}

\def\p{\partial}
\def\f{\frac}

\begin{document}
\title{van der Waals criticality in AdS black holes: a phenomenological study}
\author{Krishnakanta Bhattacharya$^{a}$\footnote {\color{blue} krishnakanta@iitg.ernet.in}}
\author{Bibhas Ranjan Majhi$^{a}$\footnote {\color{blue} bibhas.majhi@iitg.ernet.in}}
\author{Saurav Samanta$^{b}$\footnote {\color{blue} srvsmnt@gmail.com}}

\affiliation{$^a$Department of Physics, Indian Institute of Technology Guwahati, Guwahati 781039, Assam, India\\
$^b$Department of Physics, Bajkul Milani Mahavidyalaya, P. O. - Kismat Bajkul, Dist. - Purba Medinipur, Pin - 721655, India
}

\date{\today}

\begin{abstract}
AdS black holes exhibit van der Waals type phase transition. In the {\it extended} phase-space formalism, the critical exponents for any spacetime metric are identical to the standard ones. Motivated by this fact, we give a general expression for the Helmholtz free energy near the critical point which correctly reproduces these exponents. The idea is similar to the Landau model which gives a phenomenological description of the usual second order phase transition. Here two main inputs are taken into account for the analysis: (a) black holes should have van der Waals like isotherms and (b) free energy can be expressed solely as a function of thermodynamic volume and horizon temperature. Resulting analysis shows that the form of Helmholtz free energy correctly encapsulates the features of Landau function. We also discuss the {\it isolated critical point} accompanied by nonstandard values of critical exponents. The whole formalism is then extended to other two criticalities, namely $Y-X$ and $T-S$ (based on the standard; i.e. non-extended phase-space), where $X$ and $Y$ are generalized force and displacement, whereas $T$ and $S$ are horizon temperature and entropy. We observe that in the former case Gibbs free energy plays the role of Landau function, whereas in the later case that role is played by the internal energy (here it is black hole mass). Our analysis shows that, although the existence of van der Waals phase transition depends on the explicit form of the black hole metric, the values of the critical exponents are universal in nature.

\end{abstract}

\maketitle

\section{Introduction}
Black hole shows considerable similarity with other thermodynamical objects. The temperature, entropy etc. associated with the event horizon, satisfy identical laws as those of the standard thermodynamics. This fact has been known for quite some time from the pioneering works of Hawking and Bekenstein \cite{Bekenstein:1973ur,Hawking:1974sw,Bardeen:1973gs}. Later this similarity has been extended upto phase transition level \cite{Davies:1978mf,Hawking:1982dh} which shows that phase structure has one to one correspondence between black holes and standard thermodynamical systems. 

In present day, though everybody agrees on the possibility of black hole phase transition (see \cite{Mandal:2016anc,Banerjee:2016nse} for complete list and a general framework in these directions), there is no universal agreement on its interpretation. In one approach it is shown that the AdS black holes can exhibit a phase transition quite similar to liquid vapour transition. This was first noticed in \cite{Chamblin:1999tk,Chamblin:1999hg} in the non-extended phase space approach. Extended phase space description was later found in \cite{Kubiznak:2012wp}. The remarkable thing is that the similarity is not only qualitative but also quantitative. The critical exponents of both the systems (liquid vapour and black hole) are exactly same (for a review and complete list of works, see \cite{Hennigar:2015wxa,Kubiznak:2016qmn}). In this approach, cosmological constant is treated as pressure, in addition to other variables -- temperature, entropy, electric charge etc. This extended phase space approach has been found to be quite successful in describing phase transition of various AdS black holes (recent progresses in this direction can be followed from \cite{Li:2016bog,Hennigar:2016gkm,Hendi:2016usw,Upadhyay:2017fiw,Bhattacharya:2017hfj}). 

In a previous paper \cite{Majhi:2016txt}, two of the authors showed that it is not necessary to calculate the critical exponents case by case for this type of black hole phase transition. If the existence of van der Waals like phase transition is assumed, no other assumption about the metric is necessary to obtain the values of the critical exponents. This naturally explains the universality of critical exponents for this type of black hole phase transition.

In thermodynamics, critical phenomena is a thoroughly analyzed topic. There it is well known that critical exponents of completely dissimilar  processes (like liquid-vapour and ferromagnet to paramagnet phase transition) can be exactly same (see \cite{Stanley1, book} for details). The reason is that, near the phase transition point, upto certain order,  van der Waals equation for fluid and Weiss equation for magnet are same. Taking this hint, Landau made a brilliant guess about a common free energy for all second order phase transition. The beauty of this approach is that, the expression of energy is obtained from general symmetry consideration without invoking any microscopic detail. This theory plays a central role in the study of critical phenomena because the values of critical exponents can be calculated easily from this Landau free energy (also known as {\it Landau function}). 

It has been observed that, in the aforementioned type of black hole phase transition, equation of state is quite similar to the van der Waals equation near the critical point \cite{Kubiznak:2012wp}. This suggests that a free energy for black hole can be proposed without detail consideration of the spacetime. In the present paper we explore this possibility. We find that, an expression for Helmholtz free energy can be given from general consideration alone where only assumption is the existence of van der Waals type phase transition. A sporadic attempt was done earlier in \cite{Chamblin:1999tk,Chamblin:1999hg} for non-extended phase-space approach.  

Recently, within the domain of van der Waals type phase transition some points known as {\it isolated} critical points have been found where critical exponents have different values \cite{Frassino:2016vww,Hennigar:2016ekz,Dolan:2014vba,Frassino:2014pha,Mann}. These values are somewhat unusual because they do not satisfy the scaling laws. In this paper we show that appropriate Helmholtz free energy can be constructed which will give the correct values of the exponents for the isolated critical points. The whole formalism is then used to study $Y-X$ \cite{Ma:2016aat} and $T-S$ \cite{Spallucci:2013osa,Shen:2005nu,Zhang:2015ova,Mo:2016apo,Zeng:2015tfj,Zeng:2015wtt,Kuang:2016caz} criticalities in non-extended phase space. In the former case Gibbs free energy and in the later case internal energy (for black holes, it is the mass) serve the purpose. Reassuringly in all the cases we get correct values of the critical exponents. Thus we show critical phenomena can be described correctly by three energy functions for three different cases. (i) For $P-V$ criticality it is Helmholtz free energy, (ii) for $Y-X$ criticality it is Gibbs free energy and (iii) for $T-S$ criticality it is internal energy. These are the Landau functions of this type of black hole phase transition. 

The organization of our paper is as follows. Section \ref{PV} contains the $P-V$ criticality. Next two sections are devoted for analyzing the other two criticalities: $Y-X$ and $T-S$, respectively. Finally we conclude in the last section. 
\section{\label{PV}$P-V$ criticality}
     For AdS black holes, with varying cosmological constant, the first law of thermodynamics takes the form:
\begin{equation}
dM=TdS+X_idY^i+VdP~,
\label{1stlaw1}
\end{equation}
where $M$, $T$ and $S$ are mass, Hawking temperature and horizon entropy respectively. Here $X_i$ are generalized forces (e.g. electric potential $\Phi=Q/r_+$ or angular velocity $\Omega$) and $Y^i$ are generalized displacements (e.g electric charge $Q$ or angular momentum $J$). In the above, cosmological constant has been interpreted as pressure ($P= -\Lambda/8\pi$) and $V$ is known to the thermodynamic volume (for details of the above equation and concept of thermodynamic volume, see \cite{Bhattacharya:2017hfj}). Note that there is a sum over index $i$ (with Einstein's summation convention) which runs over all possible charges of the black hole. 
It has been observed that Eq. (\ref{1stlaw1}) can be cast to the standard form of the first law of black hole mechanics 
\begin{equation}
dE = TdS +X_idY^i-PdV~,
\label{1stlaw2}
\end{equation}
with energy identified as $E=M-PV$. So here mass $M$ is treated as enthalpy.

      The Helmholtz free energy $F$ is defined as
\begin{equation}
F=E-TS~.
\label{Helmohtz}
\end{equation}
Therefore the variation of the free energy, with the help of (\ref{1stlaw2}), can be written as
\begin{equation}
dF=-SdT-PdV+X_idY^i~.
\label{dF}
\end{equation}
Therefore, in general $F$ is a function of independent variables $T$, $V$ and $Y^i$. It has been observed that AdS black holes show van der Waals type feature when $Y^i$s are kept constant \cite{Kubiznak:2012wp}. Since we are interested in this case, we assume that there exists this type of phase transition and hence one can consider $F=F(V,T)$. Under this circumstance, Eq. (\ref{dF}) implies entropy and pressure can be obtained from free energy by the following relations:
\begin{equation}
S = -\Big(\frac{\partial F}{\partial T}\Big)_V; \,\,\,\  P = -\Big(\frac{\partial F}{\partial V}\Big)_T~.
\label{SP}
\end{equation}
It must be noted that, critical point is determined by the point of inflection of the isotherm. This means two conditions must be satisfied: $(\partial P/\partial V)_T|_c=0=(\partial^2P/\partial V^2)_T|_c$. Here $c$ refers to the critical point. These, by the second relation of Eq. (\ref{SP}), turn out to be
\begin{eqnarray}
&&\Big(\frac{\partial P}{\partial V}\Big)_T\Big|_c = - \Big(\frac{\partial^2F}{\partial V^2}\Big)_T\Big|_c=0~;
\nonumber
\\
&&\Big(\frac{\partial^2P}{\partial V^2}\Big)_T\Big|_c = - \Big(\frac{\partial^3F}{\partial V^3}\Big)_T\Big|_c=0~.
\label{FF}
\end{eqnarray}
These will be useful for the next calculation.

The thermodynamic quantities which show singular behaviour on the critical point are related by the critical exponents ($\alpha, \beta, \gamma, \delta$) as \cite{Cai:2013qga} {\footnote{Usually the critical exponents are defined in terms of specific volume. Since for AdS black holes, specific volume is proportional to the thermodynamic volume $V$, we introduce the definitions in terms of $V$ to keep the analysis simple.}}:
\begin{eqnarray}
&& (P-P_c) \sim (V-V_c)^{\d}~;
\nonumber
\\
&& (V_g-V_l) \sim (T-T_c)^{\b}~;
\nonumber
\\
&& C_V \sim (T-T_c)^{-\a}~;
\nonumber
\\
&& K_T \sim (T-T_c)^{-\gamma}~.
\end{eqnarray} 
In the above $C_V$ and $K_T$ are the specific heat at constant volume and isothermal compressibility, respectively. 
Since we are interested on the behavior of the AdS black holes near the critical point, let us expand the Helmholtz free energy around this:   
\begin{eqnarray}
&& F(V,T)= a_{00}+a_{10}(V-V_c)+a_{01}(T-T_c)
\nonumber
\\
&& +a_{11}(V-V_c)(T-T_c)+a_{02}(T-T_c)^2 
\nonumber
\\
&&+a_{21}(V-V_c)^2(T-T_c)+a_{12}(V-V_c)(T-T_c)^2
\nonumber
\\
&&+a_{03}(T-T_c)^3+a_{40}(V-V_c)^4+a_{31}(V-V_c)^3(T-T_c)
\nonumber
\\
&&+a_{22}(V-V_c)^2(T-T_c)^2+a_{13}(V-V_c)(T-T_c)^3
\nonumber
\\
&&+a_{04}(T-T_c)^4+\dots~.
\label{FC1}
\end{eqnarray}
In the above, $a_{ij}$ is value of $\frac{\p^{i+j}F}{\p V^i\p T^j}$ at the critical point. Note that, coefficients $a_{20}$ and $a_{30}$ are absent due to the conditions (\ref{FF}).
We shall observe that retaining terms upto $\mathcal{O}(T-T_c)^2$ in ($T-T_c$), $\mathcal{O}(V-V_c)^4$ in ($V-V_c$) and $\mathcal{O}(V-V_c)^2(T-T_c)$ will be sufficient for determining the critical exponents. This implies, as will be explained later, the free energy is sensitive upto order $(T-T_c)^2$ as $(V-V_c) \sim (T-T_c)^{1/2}$. The same also happens for the usual van der Waals gas system (see the discussion in ``critical phenomena in fluids" of \cite{book}). In this case our free energy takes the following form:
\begin{eqnarray}
F &=& a_{00}+a_{10}'v+a'_{01}t+a'_{11}v t+a'_{02}t^2 
\nonumber
\\
&+& a'_{21}v^2t+a_{40}'v^4~,
\label{Ffinal}
\end{eqnarray}
where we have used
\begin{eqnarray}
t=\frac{T}{T_c}-1; \,\,\
v=\frac{V}{V_c}-1~.
\end{eqnarray}
In the above constant coefficients are rescaled quantities of those appeared in (\ref{FC1}) by the factors like $V_c$ and $T_c$. Since explicit form of these are not important for our present discussion, we do not bother about their actual values.

  With this near critical expression of $F$, let us now find the critical exponents of the theory. For that, let us note
\begin{eqnarray}
\left(\f{\p F}{\p v}\right)_t=a'_{10}+a'_{11}t+2a'_{21}vt+4a_{40}'v^3
\end{eqnarray}
Since, the thermodynamic pressure can be defined from the Helmholtz free energy as $P= - (\p F/\p V)_T= - (1/V_c) \left(\p F/\p v\right)_t$ (see second equation of (\ref{SP})), one obtains the expression of the pressure in the following form
\begin{eqnarray}
P=P_c+a''_{11}t+a_{21}''vt+a_{40}''v^3~, 
\label{PRESS}
\end{eqnarray}
where the double primed constants are again rescaled values of the earlier constants. This notation will be adopted in later analysis also and hence from now on we shall not mention this explicitly.
Defining $\pi=P/P_c$, one gets Eq. \eqref{PRESS} in a more useful form,
\begin{eqnarray}
\pi=1+a'''_{11}t+a_{21}'''vt+a_{40}'''v^3~. \label{PI}
\end{eqnarray}
For the isothermal curves (fixed $t$) in the P-V diagram, it has to satisfy Maxwell's construction $\oint vdp=0$ \cite{book} (as it has been already assumed that the AdS black hole perform van der Waals type phase transition), where $p=P/P_c-1=\pi-1$. Now note that $dp=d\pi=(a_{21}'''t+3a_{40}'''v^2)dv$, giving
\begin{eqnarray}
\int_{v_l}^{v_g}v(a_{21}'''t+3a_{40}'''v^2)dv=0~, \label{MAX}
\end{eqnarray}
where $v_l$ and $v_g$ are the volumes of the black hole at two phases.
The above equation is valid for all the (non-critical) isotherms, which are near the critical point. Let us now focus on those isotherms which are very near to the critical one. If one considers a particular non-critical isotherm, then $t$ is constant and $v\rightarrow 0$ for the curve and therefore, the second term in \eqref{MAX} can be neglected in those cases. Thereafter, performing the integration and equating it to zero one gets $v_g=-v_l$. Remember, here $v_g$ and $v_l$ are calculated from the critical point, i.e., from $v=0$.

Now since pressure is same for both $v_l$ and $v_g$, one can express \eqref{PI} in terms of $v_l$ and $v_g$, which yields
\begin{eqnarray}
\pi=1+a'''_{11}t+a_{21}'''v_gt+a_{40}'''v_g^3~; \nonumber \\
\pi=1+a'''_{11}t+a_{21}'''v_lt+a_{40}'''v_l^3~. \label{PII}
\end{eqnarray}
Since $v_g=-v_l$, writing $v_l$ in terms of $v_g$ and then subtracting the above two equations, one obtains
\begin{eqnarray}
a_{21}'''v_gt+a_{40}'''v_g^3=0~. \label{PUN}
\end{eqnarray}
Since $v_g\neq 0$, one obtains $v_g\sim(T-T_c)^{1/2}$, which implies
\begin{eqnarray}
(V_g-V_l)\sim (T-T_c)^{1/2}~.
\end{eqnarray}
So we find $\beta=1/2$. This also implies that $v\sim (T-T_c)^{1/2}$ which was mentioned earlier to write Eq. (\ref{Ffinal}).
Setting $t=0$ in the expression of $\pi$ in \eqref{PI} yields $\pi-1=c_{2}''v^3$, which implies
\begin{align}
(P-P_c)\sim (V-V_c)^{3}~,
\end{align}
and therefore we obtain $\delta=3$.

Let us now find out the relation between specific heat at constant volume and temperature.  By definition $C_V = T(\partial S/\partial T)_V$ and therefore using the first equation of (\ref{SP}) one obtains $C_V=-T\left(\p^2 F/\p T^2\right)_V=-(T/T_c^2)\left(\p^2 F/\p t^2\right)_v$. From \eqref{Ffinal}, one can straightforwardly obtain $C_V\sim (1+t)a'_{02}$. This implies that near the critical point it is just a constant; i.e.
\begin{eqnarray}
C_V \sim {\textrm{const}}~,
\end{eqnarray}
and hence $\alpha=0$.
The isothermal compressibility is defined as $K_T=-(1/V)(\p P/\p V)^{-1}_T$. From \eqref{PRESS} one obtains
\begin{eqnarray}
K_T\sim (T-T_c)^{-1}~.
\end{eqnarray}
This means $\g=1$.

Note that, the obtained values of the critical exponents i.e., $\a=0$, $\b=1/2$, $\g=1$ and $\d=3$ obeys the scaling laws of the ordinary thermodynamic systems. These are:
\begin{eqnarray}
&& \a+2\b+\g=2~,
\no 
\\
&& \a+\b(\d+1)=2~,
\no 
\\
&& \g(\d+1)=(2-\a)(\d-1)~,
\no 
\\
&& \g=\b(\d-1)~.
\label{SCALING}
\end{eqnarray}

It must be mentioned that, the values of critical exponents were obtained for various AdS black holes by considering the explicit form of the metrics. Here we see that if any black hole metric exhibit the van der Waals like isotherms and the free energy near critical point has the form (\ref{FF}), the phase transition must be associated with the above exponents.

   Before proceeding further, let us discuss about the specific choice of the free energy (\ref{Ffinal}) and the derived expression for pressure (\ref{PRESS}). It must be noted that, the expressions are valid very near to the critical point which is defined by (\ref{FF}). These conditions are based on the following fact. The isotherms; i.e. the $P-V$ diagram for constant temperature, for an AdS black hole are exactly identical to those for the standard van der Waals system. Therefore, below the critical temperature, for a non-critical isotherm (i.e. which corresponds to nonzero $(T-T_c)$), volume has three roots. As we increase the temperature, all these roots marge together to form a point of inflection, which is the critical point. Therefore, pressure should be a cubic function of volume and hence the Taylor series expansion of $F$  must terminate to $V^4$. Of course some other terms are absent due to the conditions (\ref{FF}). All these informations have been incorporated to propose the free energy in (\ref{Ffinal}). So if equation of state of a system has these properties, it should have a Helmholtz free energy of the form (\ref{Ffinal}), and consequently have the obtained critical exponents. 
   
   In Landau's original formulation it has been argued that, in a second order phase transition symmetry is spontaneously broken. Mathematically speaking, Landau energy has one minima above the critical temperature and two minima below it. Since there must be one maxima between two minima, in general, slope of Landau energy must be zero at three different points. Or Landau energy should be quartic function of the response function (volume for liquid--vapour transition or magnetization for ferromagnet--paramagnet transition)\footnote{If in the expression of Landau energy, fourth power of response function is replaced by sixth power, resulting theory turns out to be inconsistent on thermodynamic ground \cite{Stanley1}. }.

    Thus it is clear that, presence of $a'_{40}$ in (\ref{Ffinal}) is necessary due to the existence of three real roots of volume for a isotherm. Vanishing of it does not lead to van der Waals like system. The Helmholtz free energy for  AdS black holes (as far as we know) have this crucial term and hence behave as van der Waals system. In summary, to test if $F$ has similar form as (\ref{Ffinal}) we need explicit form of the black hole but once it is confirmed that they are similar to van der Waals, one does not need to consider the explicit form of metric to find the critical exponents. Those can be obtained by the general expression (\ref{Ffinal}) for free energy.

    Another point may be worth mentioning. Here everything has been obtained just from the information of the Helmholtz free energy $F$. The similar was also done in the original work \cite{Kubiznak:2012wp} for charged AdS black hole where $F$ was found out by calculating the Gibb's free energy. Using $F$, entropy was found out and consequently authors obtained the critical exponent $\alpha=0$. Here we show that, all the critical exponents can be derived just from the near critical expansion of $F$. In this sense, we are working with the same ensemble as the original one. To be specific, since throughout the analysis the charges (i.e. $Y_i$) have been taken as constant, the ensemble, here and in the original case, is the canonical one.    

Having obtained the standard values of the critical exponents for AdS black holes, let us now concentrate on an exceptional case. There are few works \cite{Mann, Frassino:2014pha, Dolan:2014vba, Frassino:2016vww, Hennigar:2016ekz} that describes the {\it isolated} critical point in Lovelock gravity characterized by a different set of critical exponents. Most of these works find the isolated critical point in 3rd-order Lovelock gravity . Presence of isolated critical point for any order Lovelock gravity has been, however, discussed in \cite{Dolan:2014vba}.  Moreover, the critical exponents, which are unlike to that of the van der Waals system, do not obey all the scaling laws mentioned in Eq. \eqref{SCALING}.

On the isolated critical point, the $P-V$ curves of various isotherms coincide (but do not cross each other); i.e. they just touch each other. Therefore, for the critical isotherm, the conditions of coincidence at the isolated singular point are given by 
\begin{eqnarray}
&& \Big(\frac{\p P}{\p T}\Big)_V\Big|_c=0~,
\no 
\\
&& \Big(\frac{\p}{\p T}\Big(\frac{\p P}{\p V}\Big)_T\Big)_V\Big|_c=0~. 
\label{CREX}
\end{eqnarray}
 Above two relations come from the fact that pressure and the slope of the $P-V$ curves (i.e., $(\p P/\p V)_T$) are same on the isolated critical point for the different isotherms; or in other words, they are independent of the temperature. Note, that isolated critical point is a thermodynamic singular point as well \cite{Frassino:2014pha} and therefore $(\p P/\p V)_T=(\p P/\p T)_V=0$. So in this case, in addition to the conditions (\ref{FF}), isotherms must satisfy (\ref{CREX}) at the critical point.
 
 As mentions above, the isolated critical point is obtained by the four conditions mentioned in \eqref{FF} and \eqref{CREX}. Therefore, the free energy can be expanded in the following form
 \begin{align}
 F =a_{00}+a_{10}'v+a'_{01}t+a'_{02}t^2+a'_{12}v t^2 +a'_{03}t^3
 \no 
 \\
 +a'_{22}v^2t^2+a_{31}'v^3t+a_{13}'vt^3+a_{40}'v^4+a_{04}'t^4~.
\label{Ffinaliso}
 \end{align}
 Due to the four conditions of the isolated critical point, the term containing $v^2, v^3, vt$ and $v^2t$ do not appear here. As we shall show later, in this case $v\sim t$ and so the terms upto $\mathcal{O}(v^4)$ are kept in the above which are sufficient to extract the required information.
In a straightforward manner, one obtains the expansion of the pressure near the isolated critical point as
 \begin{align}
p=a'''_{12}t^2+a_{22}'''vt^2+a_{31}'''v^2t+a_{13}'''t^3+a_{40}'''v^3~, \label{PRESSISO}
 \end{align}
 Note, that in this case the equal area law for Maxwell construction is redefined as 
 \begin{align}
 \int_{v_l}^{v_g} v(a_{22}'''t^2+2a_{31}'''vt+3a_{40}'''v^2)dv=0~.
 \end{align}
  Applying the same argument as earlier, one concludes $v_l=-v_g$. Also, the Eq. \eqref{PUN} gets modified as
  \begin{align}
  a_{22}'''v_gt^2+a_{40}'''v_g^3=0~,
  \end{align}
  which implies
  \begin{align}
  (V_g-V_l)\sim (T-T_c)~.
  \end{align}
  Therefore one finds $\tilde{\b}=1$.
  
  The relation between the pressure and the volume can be obtained by setting $t=0$ in Eq. \eqref{PRESSISO}, which yields
  \begin{align}
  (P-P_c)=(V-V_c)^3~,
  \end{align}
  implying $\tilde{\d}=3$ as in the earlier case.
  In this case, the variation of the specific heat with the temperature is the same as earlier, i.e. $C_V\sim (1+t)a'_{02}$. Thus
  \begin{eqnarray}
C_V \sim {\textrm{const}}~,
\end{eqnarray}
near the critical point and hence $\tilde{\alpha}=0$.
Also, the isothermal compressibility $K_T=-(1/V)(\p P/\p V)^{-1}_T$ can be obtained from \eqref{PRESSISO}, which yields
\begin{eqnarray}
K_T\sim (T-T_c)^{-2}~,
\end{eqnarray}
at $V=V_c$. This means $\tilde{\g}=2$ in this case.

Note, for the isolated critical point ($\tilde{\a}=0$, $\tilde{\b}=1$, $\tilde{\g}=2$ and $\tilde{\d}=3$) all the scaling laws of \eqref{SCALING} are not satisfied. We found that only $\tilde{\g}=\tilde{\b}(\tilde{\d}-1)$ remains valid. It implies that out of this three critical exponents (i.e., $\tilde{\b}$, $\tilde{\g}$ and $\tilde{\d}$) only two are independent.
\section{$Y-X$ criticality}
It is well known that, $Y-X$ \cite{Ma:2016aat} and $T-S$ \cite{Spallucci:2013osa,Shen:2005nu,Zhang:2015ova,Mo:2016apo,Zeng:2015tfj,Zeng:2015wtt,Kuang:2016caz} criticalities take place in the non-extended phase space, where the cosmological constant is treated as a true constant of the theory. In this section, we shall find out the exponents of the thermodynamic quantities which become singular at the critical point. Those quantities are supposed to follow the following relations in terms of the critical exponents \cite{Shen:2005nu, Wu:2000id}.
\begin{eqnarray}
&& (Y-Y_c) \sim (X-X_c)^{\d}~;
\nonumber
\\
&& (X-X_c) \sim (T-T_c)^{\b}~;
\nonumber
\\
&& C_Y \sim (T-T_c)^{-\a}~;
\nonumber
\\
&& K_T \sim (T-T_c)^{-\gamma}~.
\end{eqnarray} 
The Gibbs free energy is defined as $G=E-TS-X_iY^i$ in the present context \cite{Beauchesne:2012qk} and the internal energy $E$ is the mass of the black hole  ($M$). Using the first law, 
\begin{equation}
dE=TdS+X_idY^i~, 
\label{ESY}
\end{equation}
one obtains 
\begin{align}
dG=-SdT-Y^idX_i~. \label{GIBBS}
\end{align}
Hence, $G$ is a function of $T$ and $X_i$, i.e., $G=G(T, X_i)$; and the conjugate quantities can be found out as
\begin{align}
S=-\Big(\frac{\p G}{\p T}\Big)_{X_i}; \,\,\,\,\,\,\ Y^i=-\Big(\frac{\p G}{\p X_i}\Big)_{T}~.
\end{align}
For the sake of generality in the analysis, we have started with the presence of multiple charges in the theory. However, when one discusses the criticality in the picture of a particular charge $Y$ and its corresponding potential $X$, the other charges and the corresponding potentials are kept constant. Therefore, from here on we shall keep only $X$ and $Y$, while the other charges will be suppressed. That will not create any loss of generality. Now, the criticality conditions of the $Y-X$ criticality are written as 
\begin{eqnarray}
&&\Big(\frac{\partial Y}{\partial X}\Big)_T\Big|_c = - \Big(\frac{\partial^2G}{\partial X^2}\Big)_T\Big|_c=0~;
\nonumber
\\
&&\Big(\frac{\partial^2Y}{\partial X^2}\Big)_T\Big|_c = - \Big(\frac{\partial^3G}{\partial X^3}\Big)_T\Big|_c=0~.
\label{COND2}
\end{eqnarray}
To provide a general prescription to determine the exponents for this criticality, we expand the Gibbs free energy about the critical point:
\begin{eqnarray}
&& G(T, X)= b_{00}+b_{10}(T-T_c)+b_{01}(X-X_c)
\nonumber
\\
&& +b_{11}(T-T_c)(X-X_c)+b_{20}(T-T_c)^2 
\nonumber
\\
&&+b_{21}(T-T_c)^2(X-X_c)+b_{12}(T-T_c)(X-X_c)^2
\nonumber
\\
&&+b_{30}(T-T_c)^3+b_{31}(T-T_c)^3(X-X_c)
\nonumber
\\
&&+b_{22}(T-T_c)^2(X-X_c)^2+b_{13}(T-T_c)(X-X_c)^3
\nonumber
\\
&&+b_{04}(X-X_c)^4+b_{40}(T-T_c)^4\dots~.
\label{GCR}
\end{eqnarray}
Following the same arguments as after Eq. \eqref{FC1}, here also the coefficients $b_{02}$ and $b_{03}$ are absent in the above expansion due to the criticality conditions. 

 It will be shown later that near the critical point, $(X-X_c)\sim (T-T_c)^{1/3}$. So, in the expansion of $G(T, X)$, keeping $\mathcal{O}(T-T_c)$ in $(T-T_c)$ and $\mathcal{O}(X-X_c)^4$ in $(X-X_c)$ will be enough for our purpose. One can, however, keep the higher order terms which will not make impact in the main analysis. Therefore, keeping the relevant terms in the expansion of $G$ in terms of $t$ and $x=\frac{X}{X_c}-1$, one obtains
 \begin{align}
 G=b_{00}+b'_{10}t+ b'_{01}x+b'_{11}tx+b'_{04}x^4~.
 \label{Gfinal}
 \end{align}
 Now, $Y=(\p G/\p X)_T=(1/X_c)(\p G/\p x)_t$, which implies
 \begin{align}
 Y=Y_c+b''_{11}t+b''_{04}x^3~, \label{Y}
 \end{align}
 and so
 \begin{align}
 (Y-Y_c)\sim (X-X_c)^3~,
 \end{align}
for $T=T_c$.
  Therefore, we get $\d=3~$.
 
 Also, from \eqref{Y} we get
 \begin{align}
 (X-X_c)\sim (T-T_c)^{1/3} \,\, {\textrm{when}} \,\, Y=Y_c~,
 \end{align}
 which implies $\b=1/3$.
 
 Now, the entropy is given as $S=-(\p G/\p T)_X=-(1/T_c)(\p G/\p t)_x$, which is found out to be 
 \begin{align}
 S=b'''_{10}+b'''_{11}x~, \ \ \ \ \ \ \ \ \ \ \ \ \ \ \ \ \ \ \ \ \ \ \ \
 \no 
 \\
 =b'''_{10}+b'''_{11}\Big(\frac{Y-Y_c-b''_{11}t}{b''_{04}}\Big)^{1/3}~,
 \label{S}
 \end{align}
 To obtain the last step, we have used Eq. \eqref{Y}. It is necessary because we want to obtain the expression of the specific heat $C_Y$ using the relation $C_Y=T(\p S/\p T)_Y=(1+t)(\p S/\p t)_Y$ from \eqref{S}. Ultimately, one can get $C_Y=(1+t)t^{-2/3}$ when $Y=Y_c$. Taking the leading order contribution, we obtain 
 \begin{align}
 C_Y\sim (T-T_c)^{-2/3} \,\, {\textrm{when}} \,\, Y=Y_c~,
 \end{align}
 which implies $\a=2/3$.
 Now, $K_T=-(1/X)(\p X/\p Y)$. Therefore, from the Eq. \eqref{Y} one obtains at $Y=Y_c$
 \begin{align}
 K_T\sim t^{-\frac{2}{3}}~,
 \end{align}
 implying $\gamma=2/3$.
 
 Let us mention that the obtained values of the critical exponents in this picture ($\a=2/3$, $\b=1/3$, $\g=2/3$ and $\d=3$) also obeys the scaling laws mentioned in \eqref{SCALING}. These are again the same which were obtained by explicit use of black hole metric.
\section{$T-S$ criticality}
  In literature one can found a number of papers that discusses the $T-S$ criticality \cite{Spallucci:2013osa,Shen:2005nu,Zhang:2015ova,Mo:2016apo,Zeng:2015tfj,Zeng:2015wtt,Kuang:2016caz}. However, people usually do not study the critical exponents in this picture. Only  a few \cite{Kuang:2016caz} mentions that 
\begin{equation}
C_Y\sim (T-T_c)^{-\a}~,
\end{equation} 
and the value of $\a$ is expected to be the same as the $Y-X$ criticality picture. This relation can be obtained starting from the internal energy. The critical point is determined by the following conditions
\begin{equation}
\Big(\frac{\partial T}{\partial S}\Big)_Y\Big|_c=0=\Big(\frac{\partial^2T}{\partial S^2}\Big)_Y\Big|_c~.
\label{TS}
\end{equation}
Remember, the internal energy ($E$), which is actually the mass of the black hole in this picture, is a function of entropy and the charges (see Eq. (\ref{ESY})). In the following, we have considered the presence of a single charge ($Y$) to determine $C_Y$, as the other charges can be taken as constants while obtaining $C_Y$ and, hence, all other charges ($Y's$) and the corresponding potentials ($X's$) can be suppressed in the theory.  In this situation, from the first law $dE=TdS+XdY$, one obtains $E=E(S,Y)$ and 
\begin{equation}
T=(\p E/\p S)_Y;\,\,\,\ X=(\p E/\p Y)_S~.
\label{TX}
\end{equation}
  
  From the analogy of \eqref{Ffinal} and \eqref{Gfinal}, proceeding in the identical way the internal energy can be expanded near the critical point in terms of $s=S/S_c-1$ and $y=Y/Y_c-1$ as
\begin{align}
 E=c_{00}+c_{01}s+c_{10}y+c_{11}ys+c_{21}y^2s+c_{12}ys^2+c_{04}s^4~.
\end{align}
Ultimately using first equation of (\ref{TX}), one obtains
\begin{align}
t=b'_{11}y+b'_{21}y^2+b'_{12}ys+b'_{04}s^3~.
\label{Tfin}
\end{align}
Now as $C_Y=T(\p T/\p S)^{-1}_Y$, from Eq. \eqref{Tfin}, one obtains $C_Y\sim (1+t)t^{-2/3}$ when $y=0$. Taking the leading order contribution near the critical point
\begin{align}
C_Y\sim (T-T_c)^{-2/3}~,
\end{align}
which implies, $\a=2/3$. This exactly matches with the findings in \cite{Kuang:2016caz}.
{\section{Conclusions}}	
There are many similarities between black hole and usual thermodynamic systems. Recent studies show that if one interprets cosmological constant as a pressure term, black hole phase transition and liquid vapour transition are identical in nature. The critical exponents of both the systems are same and near the critical point equations of states are also same.

In usual thermodynamics, some common features of all second order phase transitions motivated Landau to construct a common energy function describing the behaviour of dissimilar systems near the critical point. This method is extremely powerful because structure of Landau energy can be given from general symmetry consideration without considering underlying statistical behaviour of molecules. 
 
 In this paper we followed the same spirit to study the black hole phase transition in extended space from energy point of view. We showed that, if one accepts the existence of van der Waals type phase transition for a black hole, it is possible to give an expression of Helmholtz free energy near the critical point. This Helmholtz energy can then be used to find the critical exponents. This naturally explains the universality of critical exponents. An interesting point is, within this framework, non-standard values of exponents for isolated critical point can be described.
 
 Black hole phase transition in non-extended space has also been studied. The suitable energy function for $Y-X$ criticality is Gibbs energy and for $T-S$ criticality it is internal energy. The appropriate expressions of both these functions were obtained from general arguments and then critical exponents were calculated. Thus our approach to obtain the values of exponents is reminiscent of Landau's original approach to explain the universality of critical phenomena.  
 
 Finally, it has to be mentioned that our whole analysis is based on two inputs: (a) the AdS black holes perform van der Waals like phase transition with a critical point and (b) black hole obeys the first law of thermodynamics so that the Landau functions can be expressed in terms of relevant variables. What we found is that these set of Landau functions correctly reproduce the known critical exponents. This implies that if any system has the above features, that must be associated with these sets of exponents near the critical point. Therefore our analysis gives a strong reason for such universality. Hope this will shed some light in the black hole phase transition paradigm. 
\vskip 4mm
{\section*{Acknowledgments}}
The research of one of the authors (BRM) is supported by a START-UP RESEARCH GRANT (No. SG/PHY/P/BRM/01) from Indian Institute of Technology
Guwahati, India. 

\end{document}